# Some variations on the standard theoretical models for the *h*-index: A comparative analysis

**C. Malesios**[1]

[1]Department of Rural Development, Democritus University of Thrace, 193 Pantazidou Str., Orestiada, Greece, e-mail: malesios@agro.duth.gr

**Abstract**
There are various mathematical models proposed in the recent literature for estimating the *h*-index through bibliometric measures, such as number of articles (P) and citations received (C). These models have been previously empirically tested assuming a mathematical model and predetermining the models' parameter values at some fixed constant. Here, by adopting a statistical modelling view I investigate alternative distributions commonly used for this type of point data. I also show that the typical assumptions for the parameters of the *h*-index mathematical models in such representations are not always realistic, with more suitable specifications being favorable. Prediction of the *h*-index is also demonstrated.

## Introduction

*Hirsch (2005)* introduced an indicator, the *h*-index, based on the distribution of citations received by a given researcher's publications. By definition:
*"A scientist has index h if h of his $N_p$ papers have at least h citations each, and the other ($N_p$ - h) papers have at most h citations each"*.

*Braun et al. (2005; 2006)* recommended a Hirsch-based index to qualify the impact of scientific journals. Specifically, a journal has an *h* as journal *h*-index if the journal has published *h* papers, receiving each one at least *h* citations. The *h*-index is used more and more as an indicator for the evaluation of journals (*Jokic, 2009; Malesios and Abas, 2012; Malesios and Arabatzis, 2012*). There are also various modifications of the *h*-index proposed in the literature (see, e.g., *Rousseau, 2007; Barendse, 2007; Molinari and Molinari 2008*).

Despite the ease of calculation of the *h*-index knowing the distribution of received citations C, the attempts for a concise interpretation of the theoretical properties of the *h*-index has driven researchers to investigate the dependence of the index on the basic parameters of the citation distribution (number of articles P and citations received C) and the ways that this dependence arises through mathematical functions. Hence, various theoretical models for the *h*-index based on P and C have been proposed in the literature, with three being the main representatives (see *Ye, 2009; 2011*). These are the Hirsch model (*Hirsch, 2005*), the Egghe-Rousseau model (*Egghe and Rousseau, 2006*) and the Glänzel-Schubert model (*Schubert and Glänzel, 2007*), illustrated below (Table 1). For my investigation I include, additionally to the three standard models, a two-parameter specification for the Hirsch model, suggested by *Ye (2011)*. *Ye (2011)* and *Franceschini and Maisano (2011)* point out that these models are more suitable for aggregated levels of citation data (e.g. institutions) and less suitable for single researcher data. [See also *Burrell (2013)* for an exhaustive criticism on the appropriateness of such type of models in estimating the *h*-index of a single researcher].

| model | parameters | range | reference |
|---|---|---|---|
| $h = \sqrt[a]{P}$ | $a$ | $\alpha \in (1, \infty)$ | Egghe-Rousseau model (*Egghe and Rousseau, 2006*) |
| $h = \sqrt{\dfrac{C}{\alpha}}$ | $\alpha$ | $\alpha \in (3,5)$ | Hirsch model (*Hirsch, 2005*) |
| $h = cP^{1/(\alpha+1)}(C/P)^{\alpha/(\alpha+1)}$ | $(\alpha, c)$ | $\alpha \in (1, \infty); c \in (0, \infty)$ | Glänzel-Schubert model (*Schubert and Glänzel, 2007*) |
| $h = \left(\dfrac{C}{\alpha}\right)^{1/\alpha\beta}$ | $(\alpha, \beta)$ | $\alpha \in (1, \infty); \beta = f(\alpha)$ | Two-parameter Hirsch model (*Ye, 2011*) |

***Table 1.*** *Theoretical models for the h-index based on P and C.*

*Ye (2009)* estimates journal and institutional *h*-indices based on the latter theoretical models (except for the two-parameter Hirsch model) and examines the fit of each one of those models to conclude that the Glänzel-Schubert model is best at estimating the *h*-index. The analysis - utilizing journal and institutional bibliometric data obtained from the ISI Web of Science (WoS) – is based on deterministic mathematical expressions and pre-determined values for the models' parameters. Model fit and assessment is based on visual inspection of line plots of both observed *h*-index values and predictions based on the 3 models. However, in the formulae for the *h*-index based on different assumed theoretical models for the citation distribution, there should be no presumption that the model parameter values are universal and in practice the parameters should be estimated from each data set.

In this note, a statistical modeling view is adopted instead of the deterministic mathematical expressions utilized previously, specifically I attempt to extend this type of analysis by providing parameter estimation from a Bayesian modelling perspective, in order to come up with the most realistic estimations of the parameters of interest. The Bayesian approach was chosen for certain desirable properties, such as the flexibility to fit various models of high complexity or the ability to influence the parameters of the fitted models by using prior information. Especially, the latter constitutes a desirable feature for the mathematical models examined here since the parameters are not completely unknown but restricted to certain intervals.

The mathematical models of Table 1 as originally proposed or presented by *Ye (2009)* generally imply a linear relationship between the functions of C and/or P with the journal *h*-index. For instance, *Schubert and Glänzel (2007)* empirically tests the linear regression model $E(h_i) = cP_i^{1/3}(C_i/P_i)^{2/3}$ to find a strong linear correlation between *h* and the product $P^{1/3}(C/P)^{2/3}$.

In addition to this assumption, I also fit the models using alternative distribution specifications for the mean *h*-index $E(h)$, corresponding to suitable distributions for count data, such as the Poisson and the negative binomial (NB), which is usually utilized as an overdispersed alternative of the Poisson. In this way we have the advantage of estimating the model's parameters from each specific dataset instead of using fixed predetermined values as in *Ye (2009)*.

Additionally, I assess model fit by formal model selection criteria. The results confirm merely those of *Ye (2009)*, in the sense that the results confirm model selection, however the estimated parameters do not seem to coincide with those reported previously.

**Data**

The proposed methodology is illustrated utilizing two different datasets. The journal *h*-indices, total number of publications P and associated total citations C received for journals in the fields of ecology and forestry sciences were collected from WoS (Collection date: March, 2013 and November 2011 for ecology and forestry journals respectively). All values collected refer to the entire time windows of each one of the journals included in the WoS database. Specifically, a total of 134 journals were selected from the field of ecology category, whereas 54 journals are included in the forestry category. The ISI database was chosen mainly due to that the WoS is a database generally deemed as valid and error-free by the scientometrics' community.

**Methods**

*Models*

I fit the four theoretical bibliometric models presented in Table 1 as Bayesian regression-type models, which are of the following form:

$$H_i \sim f(h_i | \theta_i)$$
$$\theta_i = h(\mu_i)$$
$$\mu_i = g(P, C)$$

where $H_i$ is the random variable of the (theoretical) *h*-index following distribution *f* (i.e. one of the Gaussian, Poisson and NB), $\theta_i = h(\cdot)$ denotes the link function of the mean *h*-index, say $\mu_i = E(h_i)$, to each one of the functions of Table 1, and finally with $g(\cdot)$ we denote each one of the four theoretical functions. Then, for each distribution we have:

$$Gaussian(\mu_i, \sigma^2): \quad h(\mu_i) = \mu_i$$

$$Poisson(\lambda_i): \quad h(\mu_i) = \log(\lambda_i)$$

$$NB(r, q_i): \quad h(\mu_i) = \frac{r(1-q_i)}{q_i}$$

With $i = (1, 2, ..., 130)$ and $i = (1, 2, ..., 54)$ for the journals in the ecology and the forestry field, respectively.

The Bayesian approach in fitting the above models makes use of the available information, which includes prior information in the form of prior distributions assigned to the model's parameters. By combining the data with prior information we obtain the posterior distribution of model parameters utilizing Markov chain Monte Carlo (McMC) sampling (*Gelman et al., 2003*). Through this approach the robustness of the models is increased – in comparison to models where the parameters are fixed – by obtaining posterior distributions and credible intervals for the parameters of interest.

*Inference*

I perform Bayesian inference using an McMC sampling scheme. Weakly-informative priors (i.e. a truncated Gaussian with zero mean and very large variance) suitably constrained to non-negative values in accordance to the range of their values as shown in Table 1 are assigned to the models' parameters, except parameter $\alpha$ of the Hirsch model constrained to the interval (3,5). The analysis was conducted using the WinBUGS statistical software (*Lunn et al., 2000*) for model fitting. Model selection was carried through the use of the posterior mean deviance $\bar{D}$ (see *Spiegelhalter et al., 2002*). Models with smaller $\bar{D}$ value are better supported by the data. All results of posterior distributions for the models' parameters have been obtained by using 5,000 iterations as initial burn-in period and an additional sample of 50,000 iterations.

**Results and Discussion**

Goodness of fit statistics for the various models are given in the following Table (Table 2). As concerns their fit, we observe that the Glänzel-Schubert model under a Gaussian distribution provided the best fit to the two sets of data, as indicated by the values of the fit statistics ($\bar{D}$ = 894.4 and 302.2 for the ecology and forestry journals, respectively). Worst fit was observed for all theoretical models under a Poisson specification. It is evident that the large variability in the data of this type constitutes the assumption of a simple model based on the Poisson distribution for the *h*-index very unreliable. This is due to the fact that the Poisson distribution has one free parameter and does not allow for the variance to be adjusted independently of the mean. The alternative for dealing with overdispersion to the data, negative binomial model seems to perform better, especially for the two-parameter Hirsch model, where the fit outperforms those of all the rest of the models, except the Glänzel-Schubert Gaussian model. The NB specification has been shown to be more suitable also for the Egghe-Rousseau model ($\bar{D}$ =1210 and 398.7).

| Model | Distribution | $\bar{D}$ (ecology) | $\bar{D}$ (forestry) |
|---|---|---|---|
| Egghe-Rousseau | Gaussian | 1272 | 446.1 |
| Hirsch | Gaussian | 1186 | 456.8 |
| Glänzel-Schubert | Gaussian | **894.4** | **302.2** |
| Two-parameter Hirsch | Gaussian | **1021** | **328.6** |
| Egghe-Rousseau | Poisson | 2768 | 609.6 |
| Hirsch | Poisson | 403300 | 75990 |
| Glänzel-Schubert | Poisson | 16460 | 2011 |
| Two-parameter Hirsch | Poisson | 16730 | 4114 |
| Egghe-Rousseau | NB | 1210 | 398.7 |
| Hirsch | NB | 2344 | 789.2 |
| Glänzel-Schubert | NB | 1566 | 500.4 |
| Two-parameter Hirsch | NB | **959.8** | **311.2** |

***Table 2.** Mean deviance ($\bar{D}$) for the fitted models (with bold indicating the three best model fits to the data).*

As concerns the theoretical models, we see that the specification of *Ye (2011)* based upon the two parameters for the Hirsch model, at least for the Gaussian and NB distributional specifications, presents itself as a useful alternative to the Glänzel-Schubert model, although based solely on the number of citations C. Figures 1 & 2 present a visual inspection of the fit of the Glänzel-Schubert Gaussian and the two-parameter NB Hirsch models. Although both models exhibit good fit, the graphs are also revealing of the superiority of the Glänzel-Schubert Gaussian model, especially in case of predictions for the higher *h*-index values.

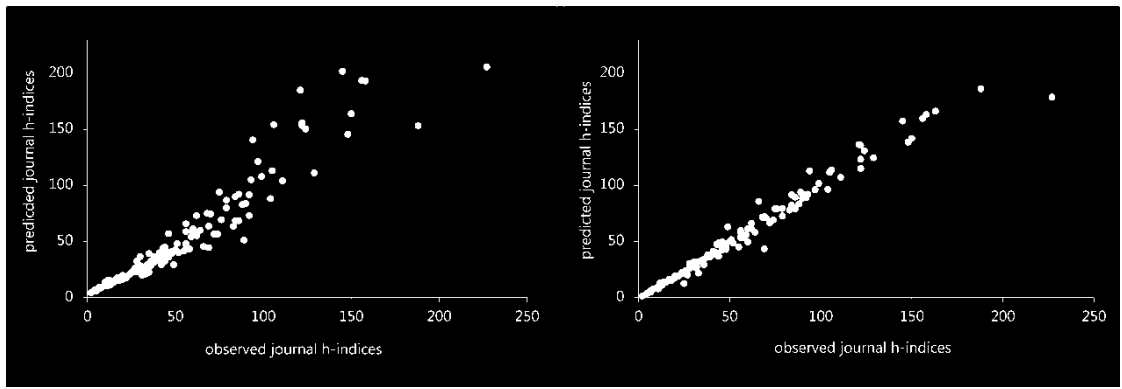

***Figure 1.** Observed vs predicted journal h-indices for the two-parameter NB Hirsch model (left) and the Glänzel-Schubert Gaussian model (right) [ecology field].*

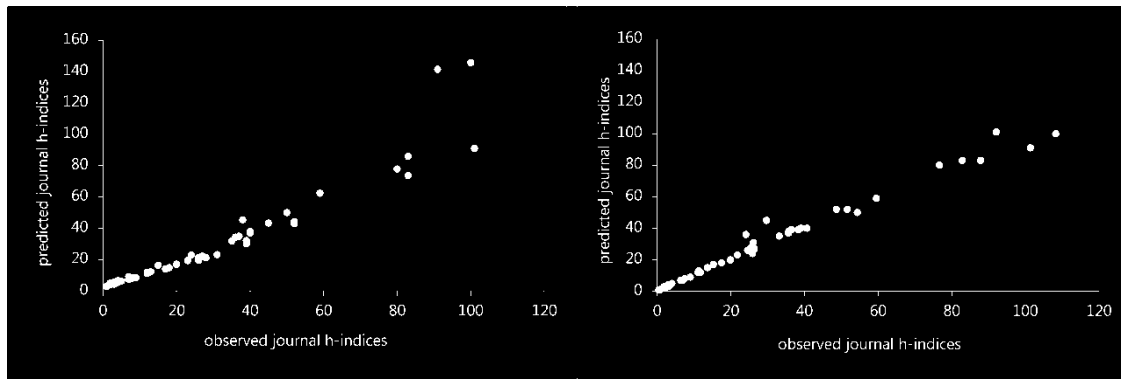

*Figure 2. Observed vs predicted journal h-indices for the two-parameter NB Hirsch model (left) and the Glänzel-Schubert Gaussian model (right) [forestry field].*

Posterior estimates results (i.e. posterior medians along with the corresponding 95% credible intervals) of parameters of interest obtained from the Bayesian models are summarized in Table A1 in the Appendix. My proposal offers credible intervals for the various parameters of the theoretical models, and compares these intervals with the already proposed values. For example, we see that my estimations for the parameter $c$ in the (Gaussian)Glänzel–Schubert model are around 0.7, coinciding thus with the empirical estimations reported in *Schubert and Glänzel (2007)* instead of the fixed value of 0.9 adopted by *Ye (2009)*. There are also substantial variations between the parameter estimates for the two fields of research, indicating thus that the model parameter values are not universal.

In summary, I have shown through an empirical statistical analysis that alternative formulations based on the three standard mathematical models for the *h*-index may result inimproved model fit. Specifically, variations related to distributional assumptions for the theoretical models and to the parameters associated with these models in certain instances resulted in better predictions. The latter were demonstrated using journal citation data from the fields of ecology and forestry.

# APPENDIX

| Model | Distribution | parameters | | | | | |
|---|---|---|---|---|---|---|---|
| | | (ecology) | | | (forestry) | | |
| | | *a* | *b* | *c* | *a* | *b* | *c* |
| Egghe-Rousseau | Gaussian | 1.811 (1.78-1.84) | -- | -- | 2.012 (1.95-2.07) | -- | -- |
| Hirsch | Gaussian | 4.985 (4.91-4.99) | -- | -- | 4.957 (4.77-4.99) | -- | -- |
| Two-parameter Hirsch | Gaussian | 3.589 (1.87-5.6) | 0.603 (0.37-1.23) | -- | 1.865 (0.83-4) | 1.284 (0.55-3.11) | -- |
| Glänzel-Schubert | Gaussian | 1.77 (1.65-1.89) | -- | 0.7 (0.64-0.75) | 1.966 (1.68-2.3) | -- | 0.784 (0.65-0.95) |
| Egghe-Rousseau | Poisson | 5.434 (5.41-5.45) | -- | -- | 5.827 (5.77-5.88) | -- | -- |
| Hirsch | Poisson | 3369 (3367-3371) | -- | -- | 1374 (1371-1376) | -- | -- |
| Two-parameter Hirsch | Poisson | 4.278 (4.14-4.35) | 2.394 (2.39-2.4) | -- | 4 (3.99-4.11) | 4.276 (4.1-4.29) | -- |
| Glänzel-Schubert | Poisson | 1.995 (1.93-2.06) | -- | 0.023 (0.022-0.024) | 3.928 (3.49-4.48) | -- | 0.084 (0.07-0.09) |
| Egghe-Rousseau | NB | 5.346 (5.25-5.44) | -- | -- | 5.798 (5.58-6.03) | -- | -- |
| Hirsch | NB | 4.926 (4.94-4.99) | -- | -- | 4.878 (4.38-4.99) | -- | -- |
| Two-parameter Hirsch | NB | 1.031 (1-1.17) | 7.11 (6.18-7.35) | -- | 1.593 (1.03-3.92) | 4.441 (1.63-7.15) | -- |
| Glänzel-Schubert | NB | 1.488 (1.11-2.07) | -- | 0.04 (0.02-0.05) | 2.011 (1.29-3.1) | -- | 0.07 (0.04-0.13) |

***Table A1.*** *Posterior parameter estimates (medians) along with the 95% credible intervals of the theoretical models for the h-index.*